\pdfoutput=1

\documentclass[11pt,twoside,a4paper,cmspaper,final,collab]{cms-tdr}
\def\svnVersion{374702}\def\cmsCernNoTag{CERN-EP/2016-161}\def\cmsCernDate{\today}\def\cmsMessage{Published in Physics Letters B as \href{http://dx.doi.org/10.1016/j.physletb.2016.11.001}{\doi{10.1016/j.physletb.2016.11.001.}}}

\begin{document}\cmsNoteHeader{BPH-13-009}

\hyphenation{had-ron-i-za-tion}
\hyphenation{cal-or-i-me-ter}
\hyphenation{de-vices}

\RCS$Revision: 370180 $
\RCS$HeadURL: svn+ssh://svn.cern.ch/reps/tdr2/papers/BPH-13-009/trunk/BPH-13-009.tex $
\RCS$Id: BPH-13-009.tex 370180 2016-10-07 17:38:08Z alverson $
\newlength\cmsFigWidth
\ifthenelse{\boolean{cms@external}}{\setlength\cmsFigWidth{0.85\columnwidth}}{\setlength\cmsFigWidth{0.4\textwidth}}
\ifthenelse{\boolean{cms@external}}{\providecommand{\cmsLeft}{top}}{\providecommand{\cmsLeft}{left}}
\ifthenelse{\boolean{cms@external}}{\providecommand{\cmsRight}{bottom}}{\providecommand{\cmsRight}{right}}

\newcommand{\BF}{\ensuremath{(4.0\pm 0.4\stat\pm 0.6\syst\pm 0.2\,(\mathcal{B}))\times 10^{-6}}\xspace}
\newcommand{\BFnosys}{\ensuremath{(4.0\pm 0.4\stat)\times 10^{-6}}\xspace}
\providecommand{\PKmp}{\ensuremath{\cmsSymbolFace{K}^\mp}\xspace}
\providecommand{\PfDz}{\ensuremath{\cmsSymbolFace{f}_0}\xspace}

\cmsNoteHeader{BPH-13-009}
\title{\texorpdfstring{Observation of the decay $\PBp\to \Pgy \Pgf \PKp$ in pp collisions at $\sqrt{s} =8$\TeV}{Observation of the decay B+ to psi(2S) phi K+ in pp collisions at sqrt(s) = 8 TeV}}

\date{\today}

\abstract{
The decay $\PBp \to \Pgy \Pgf \PKp$  is observed for the first time using data collected from pp collisions at $\sqrt{s} = 8$\TeV by the CMS experiment at the LHC, corresponding to an integrated luminosity of 19.6\fbinv. The branching fraction of this decay is measured, using the mode $\PBp\to \Pgy \PKp$ as normalization, to be \BF, where the third uncertainty is from the measured branching fraction of the normalization channel. 
}

\hypersetup{%
pdfauthor={CMS Collaboration},%
pdftitle={Observation of the decay B+ to psi(2S) phi K+ in pp collisions at sqrt(s) = 8 TeV},%
pdfsubject={CMS},%
pdfkeywords={CMS, physics, B physics, rare B decays }}

\maketitle

\section{Introduction}
\label{sec:Intro}

The large cross section for b quark production at the CERN LHC and the high luminosity of the
accelerator provide the possibility to study rare B meson decays. Recently, several
experiments have reported the likely  presence of structures in the $\JPsi\Pgf$ mass spectrum from $\PBpm
\to \JPsi  \Pgf \PKpm$
decays~\cite{Aaltonen:2009tz,Aaij:2012pz,Abazov:2013xda,Chatrchyan:2013dma,Lees:2014lra, Aaij:2016iza, Aaij:2016nsc}.
A natural extension of these results is to study the $\Pgy\Pgf\PKpm$ and the $\Pgy\Pgf$ mass spectra. As part of that investigation, we report the
first observation of the decay $\PBpm \to \Pgy \Pgf\PKpm$,
with $ \Pgy\to \PGmp\PGmm$ and $\Pgf\to \PKp\PKm$.
We measure the corresponding branching fraction using data collected at the LHC with the CMS detector in proton-proton (pp) collisions at $\sqrt{s} = 8$\TeV, corresponding to an integrated luminosity of 19.6\fbinv.
Possible contributions from nonresonant $\PKp\PKm$ and $\PfDz(980)$ states in the signal are also studied, and an upper limit is determined in the fraction of events that do not correspond to $\Pgf\to\PKp\PKm$ in the $\PBpm \to \Pgy \PKpm\PKmp\PKpm$ channel. In what follows, $\phi$ is used to represent the $\Pgf$ meson, and all results are combined in the investigation of the two charge-conjugate states.

\section{The CMS detector}
\label{sec:detector}

The central feature of the CMS apparatus is a 13\unit{m} long superconducting solenoid of 6\unit{m} internal diameter, providing a magnetic field of 3.8\unit{T}. A silicon pixel and strip tracker, a lead tungstate crystal electromagnetic calorimeter, and a brass and scintillator hadron calorimeter,
each composed of a barrel and two endcap sections, reside within the volume of the solenoid. Muons are measured in gas-ionization detectors embedded in the steel flux-return yoke outside the solenoid. Extensive forward calorimetry complements the coverage provided by the barrel and endcap detectors.

The main subdetectors used in the present analysis are the silicon tracker and the muon detection system. Muons are identified within the pseudorapidity
range $\abs{\eta}< 2.4$, using detection planes based on three technologies: drift tubes, cathode-strip chambers, and resistive-plate chambers. The silicon tracker measures charged particles within the range $\abs{\eta}< 2.5$. It consists of 66 million $100{\times}150\mum^2$ silicon pixels and more than 9 million silicon strips. For reconstructed particles with transverse momenta $1 < \pt< 10$\GeV and $\abs{\eta} < 1.4$, the track resolutions are typically 1.5\% in \pt, and the transverse and longitudinal impact parameters are in the respective ranges of 25--90 and 45--150\mum~\cite{Chatrchyan:2014fea}.

The first level of the CMS trigger system, constructed from specialized hardware processors, uses information from the calorimeters and muon detectors to select the most interesting events in a time interval of less than 4\mus. The high-level trigger (HLT) processor farm further decreases the event rate from around 100\unit{kHz} to less than 1\unit{kHz}, before data storage.
 At the HLT stage, there is full access to all event information, including tracking, and therefore selection criteria similar to those applied offline can be implemented.

A more detailed description of the CMS detector, together with a definition of its coordinate system and relevant kinematic variables, can be found in Ref.~\cite{Chatrchyan:2008zzk}.

\section{Data  and event selection}
\label{sec:data}

Monte Carlo (MC) simulated events are generated using \PYTHIA 6.4~\cite{Pythia} for particle production, \EVTGEN 1.3.0~\cite{EvtGen} for particle decays, and \GEANTfour~\cite{Geant4} for tracing the particles through a detailed model of the CMS detector. These samples are generated using the conditions appropriate for the data analyzed, including the effects of alignment, efficiency, and average number of additional pp collisions and their multiple reconstructed vertices per beam crossing (pileup).

The events in the analysis were collected with a trigger based on the invariant mass and $\pt$ of the dimuon system. The following criteria are applied in the HLT: (i) the dimuon $\pt$ is required to be greater than 4.9\GeV, (ii)~the two muons must be oppositely charged, (iii) the dimuon invariant mass is required to be in the range of 3.35--4.05\GeV, and (iv) the dimuon tracks must form a three-dimensional (3D) vertex with a $\chi^2$ probability greater than 0.5\%.

The two muons from the triggered event are required to be within 5 $\sigma_{M}^{\Pgy}$ of the nominal $\Pgy$ mass~\cite{PDG}, where the standard deviation $\sigma_{M}^{\Pgy}$ is obtained from fits to signal MC events in four bins of dimuon pseudorapidity. The value of $\sigma_{M}^{\Pgy}$ varies from 23\MeV for $\abs{\eta}<0.6$ to 45\MeV for $\abs{\eta}>1.8$.
The $\PBp \to \Pgy \phi \PKp$ candidates are reconstructed by combining three additional charged particle tracks consistent with originating from the $\Pgy $ vertex, and have a total charge of 1. These tracks are assigned the kaon mass. The $\PBp$ decay vertex is reconstructed using a kinematic fit to a common 3D vertex constraining the invariant mass of the two muons to the nominal $\Pgy$ mass.
For multiple candidates, the one with the highest $\PBp$ vertex probability is retained.  The overall efficiencies in selecting the correct candidate obtained from MC studies are 96.8\% and 99.4\% for the $\PBp \to \Pgy \phi \PKp$ and $\PBp \to \Pgy \PKp$ events, respectively. The  $\pt$ of each kaon track is required to be greater than 1\GeV. Only tracks passing the standard CMS high-purity requirements~\cite{Khachatryan:2010pw} are used.
There are two $\PKp\PKm$ combinations for the three charged kaon
tracks, and the combination with invariant mass closest to the nominal $\phi$ meson mass~\cite{PDG} is used as the $\phi$ candidate. This selection yields the correct $\PKp\PKm$ pair ($94 \pm1$)\% of the time, as determined from simulation.
The mass of the $\phi$ candidate is not constrained to its nominal value  because
the  experimental $\PKp\PKm$ mass resolution (1.3\MeV, obtained from our MC simulation) is
less than the natural width of the $\phi$ meson (4.3\MeV)~\cite{PDG}.

Additional requirements are placed on the resulting sample to optimize the sensitivity to the signal mode.
The  signal region is defined to lie within $\pm$5 $\sigma_{M}^{\PB} $ of the nominal $\PBp$ mass~\cite{PDG}, where $\sigma_{M}^{\PB}$ is
determined to be 3\MeV in a fit to  simulated signal events using a  single Gaussian function.
Five quantities are chosen to optimize a 5 standard deviation discovery Punzi figure-of-merit (FOM), defined as
$N_{S}/(5/2+ \sqrt{N_{B}})$~\cite{FOM}, where
$N_{S}$ is the number of $\PBp$ candidates in the simulated sample, and  $N_{B}$ is the
number of background candidates within $\pm 5$ $\sigma_{M}^{\PB}$ of the $\PBp$ mass peak. The background contribution is obtained from a fit to the sideband events in the $\Pgy \phi \PKp$ invariant mass spectrum, where the lower and upper sidebands are defined as 5.220--5.264 and 5.294--5.330\GeV, respectively. The five quantities used to optimize the FOM are as follows: (i)~the $\PBp$ vertex probability; (ii)~the significance of the transverse displacement, defined as the ratio of the transverse distance $L_{xy}$ of the $\PBp$ secondary vertex relative to the center of the beam spot and its uncertainty $\sigma_{L_{xy}}$, with the latter being the sum in quadrature of the uncertainty in the  transverse position of the secondary vertex and the transverse size of the beam spot; (iii)~the cosine of the pointing angle $\theta$, defined as the angle between the reconstructed $\PBp$ momentum vector and its flight direction, as determined from the vector connecting
the primary vertex~\cite{Chatrchyan:2014fea} to the $\PBp$ secondary vertex, where the primary vertex is chosen so that this angle is closest to zero; (iv)~the  \pt of the dimuon system; and (v)~the $\phi$ mass window, defined as the difference between the invariant mass of the $\PKp\PKm$ system and the mass of the $\phi$ meson~\cite{PDG}.
The selection criteria derived from the optimization procedure are shown in Table~\ref{table:Optcutvalues}. The overall efficiency of the offline signal selection is $(1.91\pm 0.01)\times 10^{-3}$.

\begin{table}
\centering
\topcaption{The selection criteria derived from the optimization procedure.}
\label{table:Optcutvalues}
\begin{tabular}{ll}
 \hline
Quantity &Requirement \\
\hline
B vertex probability &$>$10\%\\
$L_{xy}/\sigma_{L_{xy}}$  &$>$4\\
$\cos\theta$&$>$0.99\\
Dimuon \pt &$>$7\GeV \\
$\phi$ mass window &$<$8\MeV \\
\hline
\end{tabular}
\end{table}

\section{Observation of \texorpdfstring{$\PBp\to \Pgy\phi \PKp$}{B+ to psi(2S) phi K+} decay}
\label{sec:observation}

\begin{figure}[!htb]
\centering
\includegraphics[width=0.48\textwidth]{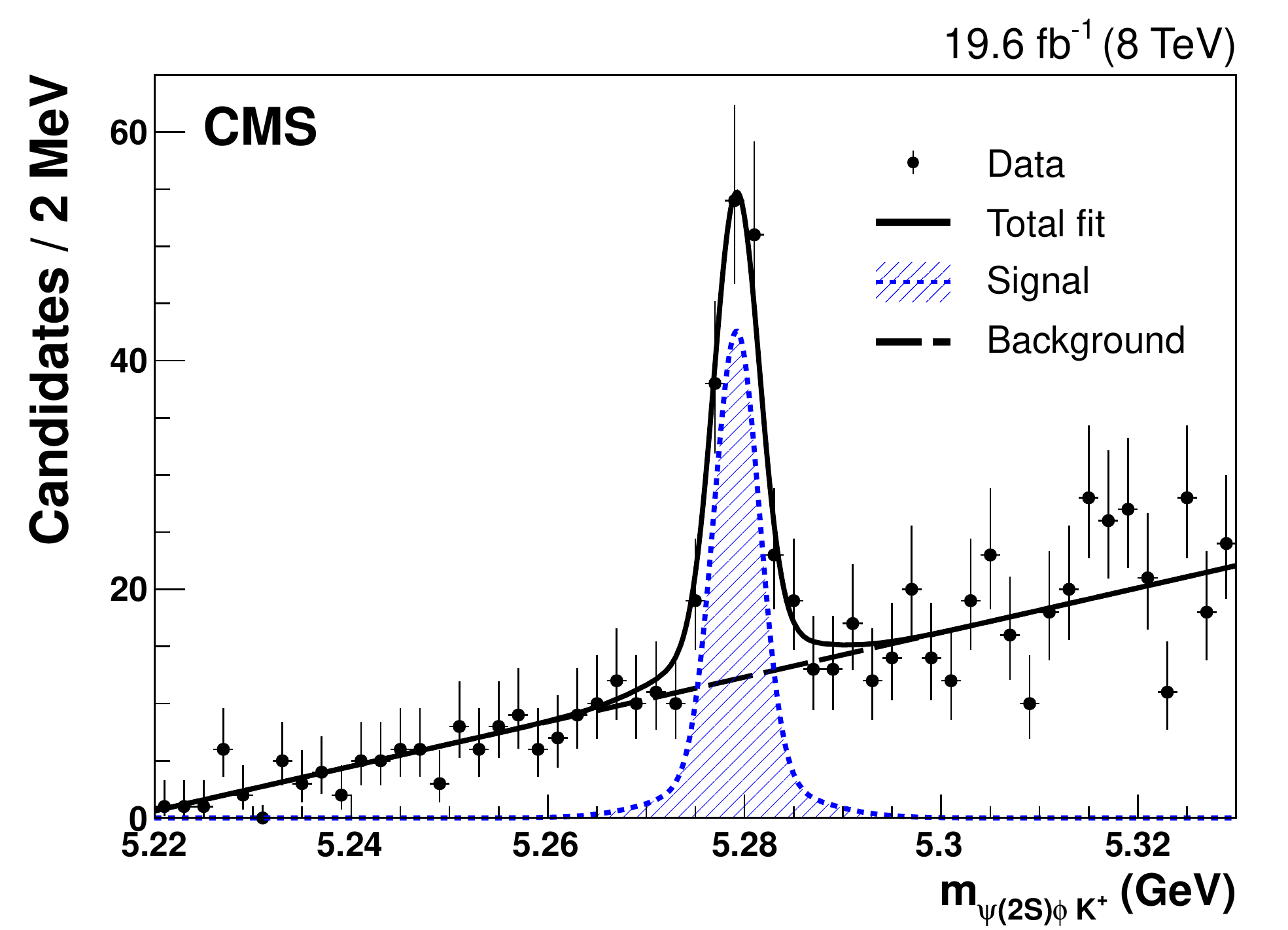}
\caption{The $\Pgy \phi \PKp$ mass distribution after the selection requirements.
The solid curve shows the result of fitting this distribution to a signal represented by two Gaussian functions and a
first-order polynomial for the background. The shaded area represents the signal component, while the black dashed line shows the fitted
background contribution.\label{fig:fig1}}
\end{figure}

The invariant mass spectrum of the selected $\Pgy\phi \PKp$ candidates is shown in Fig.~\ref{fig:fig1}. An extended unbinned maximum-likelihood estimator from RooFit~\cite{Verkerke:2003ir} is used to perform the fit to the data, using two Gaussian functions for the signal and
a first-order polynomial for the background. The two Gaussian functions share a common mean fixed to the nominal $\PBp$ mass~\cite{PDG}, while their widths and relative fractions
are fixed to the values obtained in the MC simulation. The goodness of fit is checked using a $\chi^{2}$ test, which returns a $\chi^2$ per degree of freedom (dof) of 23.0/24, with a corresponding probability of 52\%. The fit gives a $\PBp$ yield of  $140\pm15$ events, where the uncertainty is statistical.

The small amount of available phase space, only 80\MeV, in the $\PBp\to \Pgy\phi \PKp$ decay makes it difficult to model the expected $\Pgy\phi$ invariant mass distribution. This, and the limited number of signal events, precludes any search for $\Pgy\phi$ resonances in the current data sample.

Possible contamination from the decays of the $\PfDz(980)$ meson and nonresonant $\PKp\PKm$ is studied through a simultaneous fit of the $\PKp\PKm$ invariant mass distributions for the combinations closest to the nominal $\phi$ mass inside and outside of an 18\MeV mass window centered around the nominal $\PBp$ mass, not using the ${\pm}8\MeV$ $\phi$ mass window listed in Table~\ref{table:Optcutvalues}. The distributions of the nonresonant $\PKp\PKm$ and $\PfDz$ background invariant mass contributions are obtained from dedicated $\PBp \to \Pgy\PKp\PKm \PKp$ and $\PBp \to \Pgy \PfDz \PKp$ MC simulations generated using \EVTGEN~\cite{EvtGen}, which models the $\PfDz$ distribution as a coupled-channel Breit--Wigner function~\cite{Flatte:1972rz}. Both the nonresonant $\PKp\PKm$ and $\PfDz$ contributions are distorted through the selection of the $\PKp\PKm$ pair closest to the nominal $\phi$ mass. We parametrize these forms using Gaussian functions that are very similar for the two components. Their correlation coefficients show that the two contributions are strongly anticorrelated, and we therefore choose the nonresonant $\PKp\PKm$ background in the fit to represent both the $\PfDz$ and any other $\PKp\PKm$ components, and label it as the non-$\phi$ contribution to the signal.

\begin{figure*}
\centering
\includegraphics[width=0.48\textwidth]{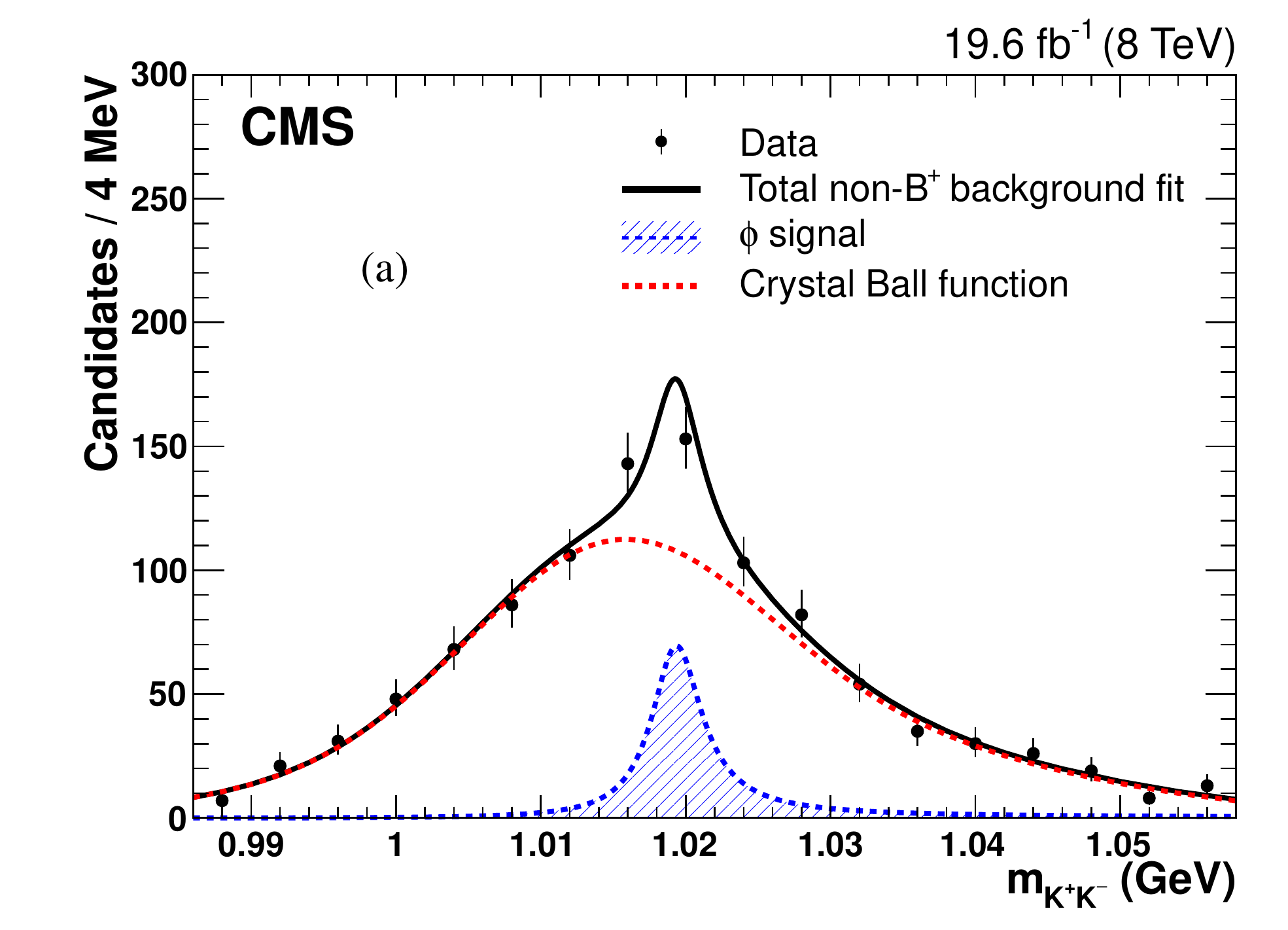}
\includegraphics[width=0.48\textwidth]{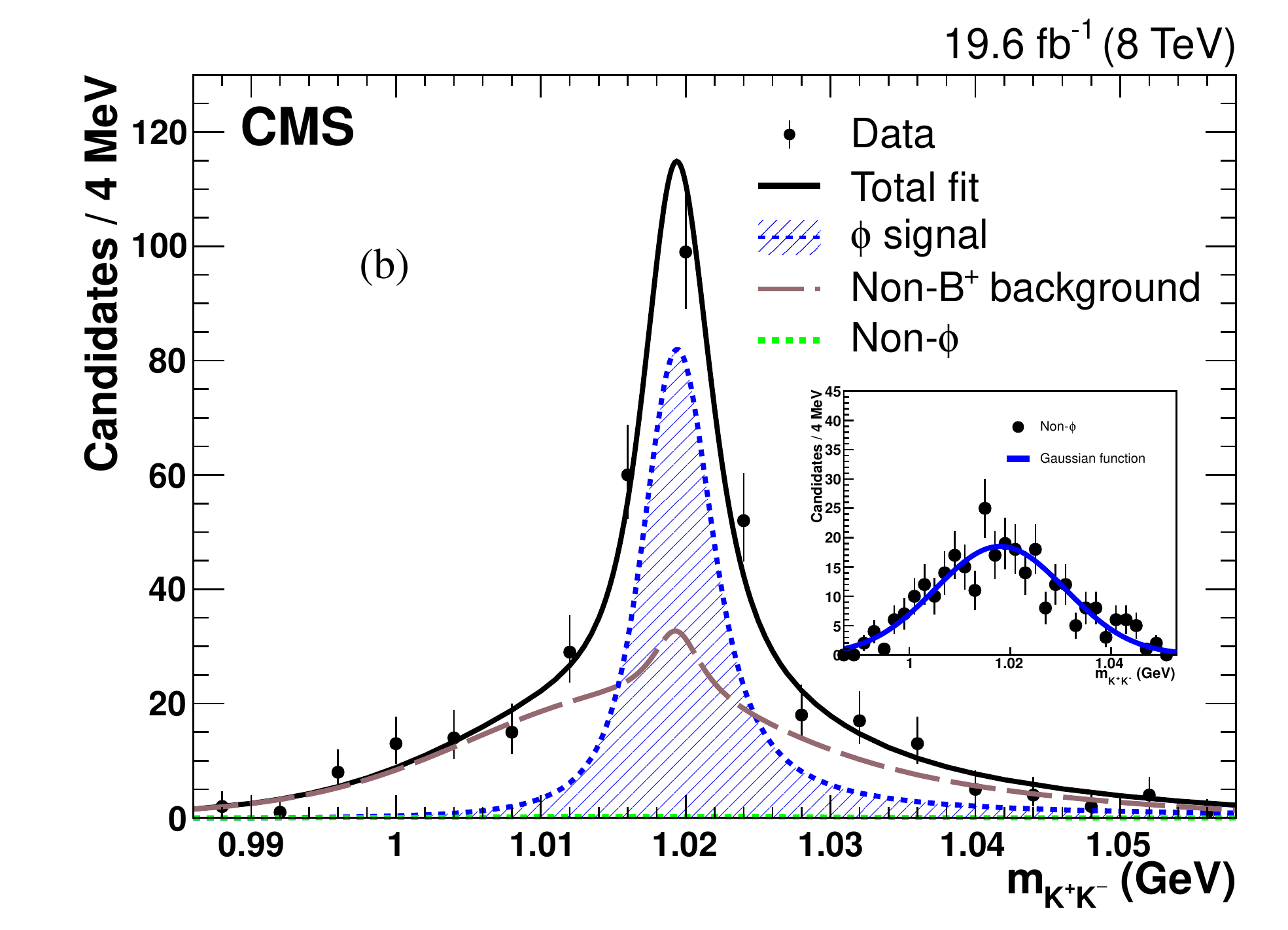}
\caption{The distributions of the $\PKp\PKm$ invariant mass closest to the nominal $\phi$ mass outside
(a) and inside (b) the $\PBp$ mass window. The results show the total fit, as well as (a) the Crystal Ball function
component of the non-$\PBp$ background, and in (b) the four components: $\phi$ signal, non-$\PBp$ background, non-$\phi$ signal. The non-$\phi$ component is almost invisible due to its small contribution. The inset shows the non-$\phi$ distribution modelled by a Gaussian function.\label{fig:simulatenousfit}}
\end{figure*}

A 3$\sigma_{M}^{\PB}$-wide window around the $\PBp$ mass of 5.270--5.288\GeV is chosen, with lower and upper sidebands of 5.220--5.270 and 5.288--5.330\GeV, respectively. Figures~\ref{fig:simulatenousfit}(a) and (b) show the distributions of the $\PKp \PKm$ invariant masses closest to the nominal $\phi$ mass outside (a) and inside (b) the $\PBp$ window, where Fig.~\ref{fig:simulatenousfit}(a) thereby provides the non-$\PBp$ background. Both the distributions share the same non-$\PBp$ background function. We find (not displayed) that the non-$\PBp$ contribution within the $\PBp$ mass window is of 194 $\pm$ 14 events, obtained from a fit to the $\Pgy\phi\PKp$ invariant mass spectrum. In the simultaneous fit, we therefore fix this number to 194, while the number of non-$\PBp$ events contributing in the sidebands is allowed to vary.

The $\phi$ signal component is parametrized by a P-wave relativistic Breit--Wigner function, convolved  with a Gaussian resolution function.
The standard deviation of the Gaussian function is fixed to 1.3\MeV. The mass and width of the $\phi$ reflect their nominal values~\cite{PDG}.
Since there is a $\phi$ signal in the non-$\PBp$ events, the non-$\PBp$ $\phi$ contribution in Fig.~\ref{fig:simulatenousfit}(a) is parametrized by the sum of a Crystal Ball function~\cite{bib:CrystalBall} and the above-mentioned function that represents the $\phi$ component.

The data in Fig.~\ref{fig:simulatenousfit}(a) is fitted using the non-$\PBp$ function, and simultaneously in Fig.~\ref{fig:simulatenousfit}(b) using the above three functions.
The fit returns a yield of $2\pm20$ events for the non-$\phi$ signal contribution that is too small to be seen in the Fig.~\ref{fig:simulatenousfit}(b). The systematic uncertainty in this yield is negligible.
The fit quality in Fig.~\ref{fig:simulatenousfit}(b) is checked using a $\chi^{2}$ test, which returns $\chi^2/\text{dof} = 21.6/16$. We set an upper limit on the fraction of the non-$\phi$ component in $\PBp \to \Pgy\PKp\PKm\PKp$ decays, obtained with the CL$_\mathrm{s}$ method~\cite{Junk:1999kv,Read:2002} using an asymptotic approximation~\cite{Asymptotic:2011},
of 0.26 at the 95\% confidence level.

\section{\texorpdfstring{$\PBp\to \Pgy\PKp$}{B+ to psi(2S) K+} decay}
\label{sec:normdecay}

\begin{figure}[!htb]
\centering
\includegraphics[width=0.48\textwidth]{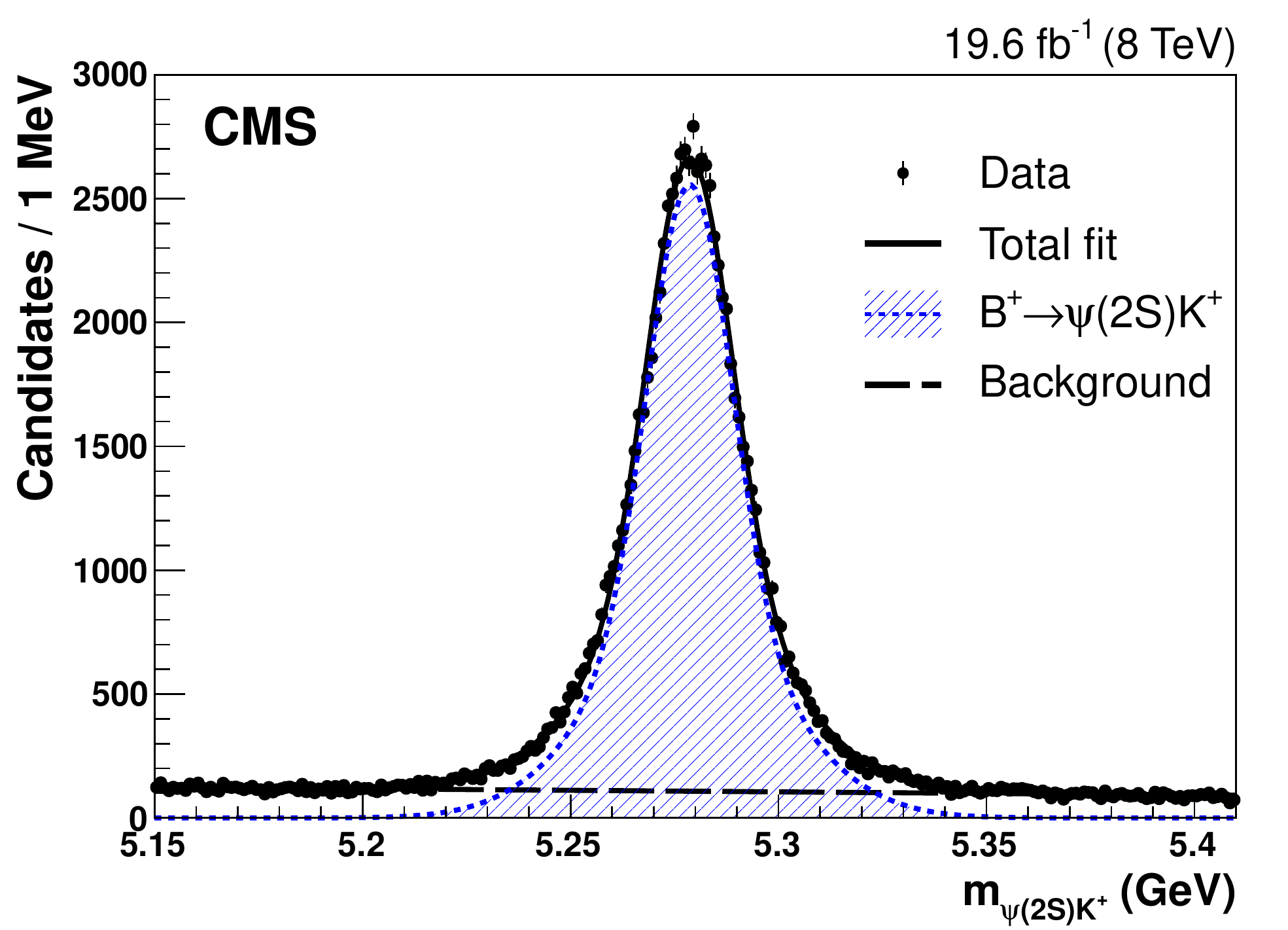}
\caption{
The $\Pgy  \PKp$ invariant mass distribution after implementing all event selection requirements.
The solid curve shows the result of fitting this distribution to a signal corresponding to two Gaussian functions and a
first-order polynomial for the background. The shaded area represents the contribution from the $\PBp\to\Pgy  \PKp$ channel, while the black dashed line shows the fitted
background contribution.\label{fig:fig2}}
\end{figure}

The $\PBp\to \Pgy\PKp$ decay is chosen as the normalization channel
because its absolute branching fraction is well measured, it is recorded with the same trigger as the signal channel,
and it is topologically similar to the signal, so that many
systematic uncertainties cancel or are reduced.
All applicable selection requirements are kept the same as those for the signal channel.
The $\Pgy\PKp$ invariant mass distribution is shown in Fig.~\ref{fig:fig2}.
A  binned maximum-likelihood fit is used to determine the number of events in this channel.
Again, two Gaussian functions are used  to model the $\PBp$ signal, and a first-order polynomial to model the background.
The large number of events allows a fit with all parameters free to vary and the yield is found to be $ 87264 \pm 363\stat$.
The goodness of the fit is checked using a $\chi^{2}$ test, which returns $\chi^2/\text{dof} = 363/253$.

\section{Results and systematic uncertainties}
\label{sec:result}

The value of the branching fraction $\mathcal{B}(\PBp\to \Pgy\phi \PKp)$ is obtained from the ratio:
\begin{equation}
\frac{N_{\PBp\to\Pgy\phi \PKp}\, \mathcal{B}(\PBp\to \Pgy\PKp)}{N_{\PBp\to\Pgy \PKp}\,\varepsilon_{\text{rel}}\, \mathcal{B}(\phi\to \PKp\PKm) },
\label{eq:brcal}
\end{equation}
where $N_{\PBp\to\Pgy\phi \PKp}$ and $N_{\PBp\to\Pgy \PKp}$ are the signal and normalization yields extracted from the fit to  the invariant mass distributions in the $\PBp\to\Pgy\phi \PKp$ and $\PBp\to\Pgy \PKp$ channels, respectively. The relative reconstruction efficiency $\varepsilon_{\text{rel}}=0.52 \pm 0.01$ is calculated from the ratio of the reconstruction efficiencies in the signal and normalization channels with a statistical uncertainty from the number of MC events.
Using the branching fractions $\mathcal{B}(\phi\to \PKp\PKm)=0.489\pm0.005$ and $\mathcal{B}(\PBp\to \Pgy\PKp)=(6.26\pm0.24)\times 10^{-4}$~\cite{PDG}, the value of $\mathcal{B}(\PBp\to \Pgy\phi\PKp)$ is determined to be $\BFnosys$.

Estimates of the contributions to the  systematic uncertainty in $\mathcal{B}(\PBp\to \Pgy \phi \PKp)$ are summarized in Table~\ref{table:Systematics}, and described below. The uncertainty from modeling the shape of the $\PBp$ invariant mass distribution
is estimated to be 8.6\% by allowing the widths of the two Gaussian functions to vary in the fit, with the background function fixed to a first order polynomial. Systematic uncertainties from sources such as muon identification, trigger efficiency, and track reconstruction efficiency for the three common tracks (two muons and a kaon) almost cancel in the measurement of the signal branching fraction. The uncertainty in the charged particle track reconstruction efficiency, obtained in an independent study by comparing two-body and four-body $\PDz$ decays in data and simulated events~\cite{Chatrchyan:2014fea}, gives an uncertainty of 3.9\% per track and a total uncertainty of 7.8\% for the two additional kaon tracks.
A mismatch in the \pt distribution between $\PBp$ mesons in MC simulations and in data can lead to an incorrect efficiency.
We therefore reweight the signal and normalization events
using a weighting function derived from the normalization channel.
The ratio of efficiencies from the reweighted MC events is compared to the nominal value to extract a systematic uncertainty of 5.3\%.

\begin{figure}[!htb]
\centering
\includegraphics[width=0.48\textwidth]{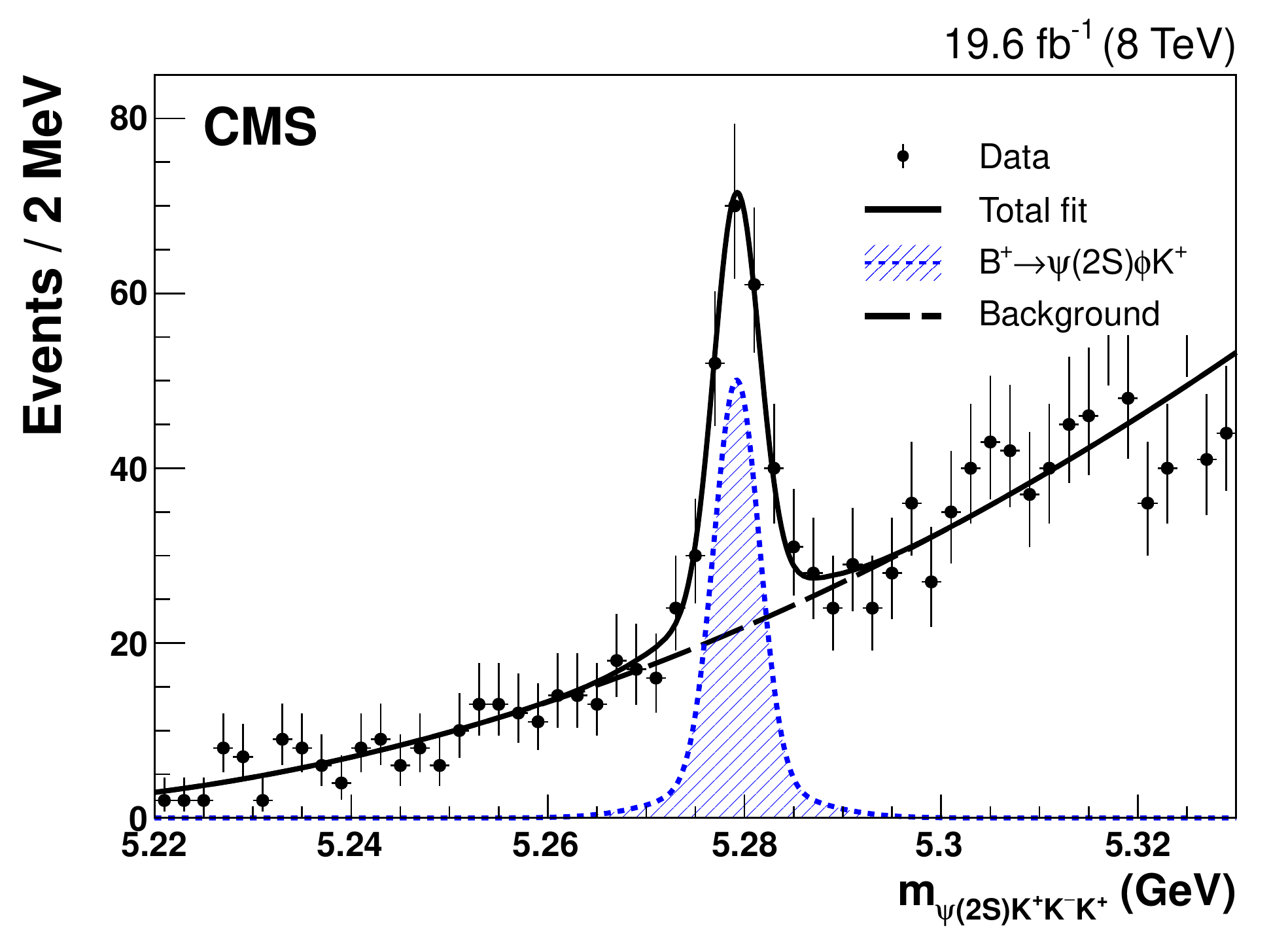}
\caption{
The $\Pgy \PKp\PKm \PKp$ invariant mass distribution  with no $\phi$ mass selection.
The solid curve shows the result of fitting this distribution to a signal represented by two Gaussian functions and a
second-order polynomial for the background. The shaded area represents the signal component, while the black dashed line shows the fitted
background contribution.    \label{fig:fig3}}
\end{figure}

 The choice of the $\PKp\PKm$ candidate closest to the nominal $\phi$ mass causes a bias, and, to estimate any systematic contamination of the $\PKp\PKm$ mass peak from non-$\phi$ backgrounds, the analysis is repeated after removing the selection on the $\PKp\PKm$ mass being closest to the mass of the $\phi$. This makes the choice of the $\PKp\PKm$ pair independent of the closest value to the nominal $\phi$ mass, and the branching fraction is remeasured by keeping both $\PKp\PKm$ pair candidate events. The subsequent $\PBp\to \Pgy \PKp\PKm\PKp$ invariant mass distribution is shown in Fig.~\ref{fig:fig3}. The signal in Fig.~\ref{fig:fig3} is clear, but there is more background relative to the signal mass distribution shown in Fig.~\ref{fig:fig1}. There are  $165\pm 18 $ $\PBp$ signal events with two $\PKp\PKm$ combinations for each event. The efficiency for the $\PBp\to \Pgy \phi \PKp$ signal after removing the choice of $\phi$ candidate is $(2.14\pm0.02)\times 10^{-3}$, and the redetermined $\mathcal{B}(\PBp\to \Pgy\phi \PKp)$ is $(4.2\pm0.4\stat)\times 10^{-6}$. The 5.0\% difference between this and the nominal branching fraction is used as the systematic uncertainty from possible non-$\phi$ backgrounds.

\begin{table*}[!htb]
\centering
\topcaption{Relative systematic uncertainties in the measurement of $\mathcal{B}(\PBp\to \Pgy \phi\PKp)$ in percent. The total systematic uncertainty corresponds to the sum in quadrature of the listed uncertainties.}
\label{table:Systematics}
\begin{tabular}{lc}
 \hline
Source &  Uncertainty (\%) \\
\hline
$\PBp$ mass shape for signal mode  &8.6\\
Charged particle track reconstruction efficiency &  7.8\\
Modeling of \pt dependence of $\PBp$ efficiency&5.3\\
$\phi$ purity&5.0\\
Mass distribution for the background in the signal  & 2.9\\
Uncertainty in relative efficiency of signal and normalization  & 2.3\\
Background distribution in the normalization channel&2.2\\
Angular distributions of $\PKp\PKm$ systems&  1.9\\
$\PBp$ mass shape for normalization mode &1.0\\
$\mathcal{B}(\phi\to\PKp\PKm)$ uncertainty &1.0\\
\hline
Total &  15\\
\hline
\end{tabular}
\end{table*}

 The uncertainties in modeling the $\PBp\to \Pgy \phi \PKp$ and the normalization channel backgrounds are estimated to be 2.9\% and 2.2\%, respectively, by adding polynomials of higher order in the fit to describe the background.
The uncertainty from the angular distribution of the $\PKp\PKm$ system is estimated to be 1.9\%, based on the changes induced in the $\PBp$ reconstruction efficiency by weighting the simulated events with different helicity angle distributions. The uncertainty in the $\PBp$ mass shape for the normalization channel is estimated to be 1.0\% by adding a third Gaussian function with a common mean and a varying width to the fit, with the background again modeled by a linear function. The uncertainty in $\mathcal{B}(\phi\to\PKp\PKm)$ is 1\%~\cite{PDG}.

 Possible systematic uncertainties introduced by different trigger and pileup conditions and analysis selections have been investigated by dividing the data into subsets and evaluating the statistical consistency~\cite{PDG} of the independent samples; the resulting variations are found to be within the expected uncertainties.

The overall relative systematic uncertainty in $\mathcal{B}(\PBp\to \Pgy\phi \PKp)$ is 15\% from adding the individual contributions summarized in Table~\ref{table:Systematics} in quadrature. The total uncertainty in $\mathcal{B}(\PBp\to \Pgy\PKp)$ is 3.8\%~\cite{PDG}. Thus, the final measured value of $\mathcal{B}(\PBp\to \Pgy\phi \PKp)$, including all systematic uncertainties, is: $\BF$, where the third uncertainty is from the imprecision in $\mathcal{B}(\PBp\to \Pgy\PKp)$.

\section{Summary}
\label{sec:conclusion}

The $\PBp\to
\Pgy\phi \PKp$ decay has been observed for
the first time by the CMS Collaboration. The result has been achieved using data from pp collisions at $\sqrt{s}$ = 8\TeV, corresponding to an integrated luminosity of 19.6\fbinv. The branching fraction
$\mathcal{B}(\PBp\to \Pgy\phi\PKp)$ is determined to be \BF, where the third
uncertainty arises from the imprecision in the value of
$\mathcal{B}(\PBp\to
\Pgy\PKp)$. The upper limit on the fraction
of $\PBp\to \Pgy
(\text{non-}\phi)\PKp$ decays in $\PBp \to \Pgy \PKp\PKp\PKp$ channel is found to be 0.26 at 95\% confidence. The observation
of $\PBp\to \Pgy\phi
\PKp$ offers future opportunities in
searches for resonances in the $\Pgy\phi$ mass spectrum.

\begin{acknowledgments}
We congratulate our colleagues in the CERN accelerator departments for the excellent performance of the LHC and thank the technical and administrative staffs at CERN and at other CMS institutes for their contributions to the success of the CMS effort. In addition, we gratefully acknowledge the computing centers and personnel of the Worldwide LHC Computing Grid for delivering so effectively the computing infrastructure essential to our analyses. Finally, we acknowledge the enduring support for the construction and operation of the LHC and the CMS detector provided by the following funding agencies: BMWFW and FWF (Austria); FNRS and FWO (Belgium); CNPq, CAPES, FAPERJ, and FAPESP (Brazil); MES (Bulgaria); CERN; CAS, MoST, and NSFC (China); COLCIENCIAS (Colombia); MSES and CSF (Croatia); RPF (Cyprus); SENESCYT (Ecuador); MoER, ERC IUT and ERDF (Estonia); Academy of Finland, MEC, and HIP (Finland); CEA and CNRS/IN2P3 (France); BMBF, DFG, and HGF (Germany); GSRT (Greece); OTKA and NIH (Hungary); DAE and DST (India); IPM (Iran); SFI (Ireland); INFN (Italy); MSIP and NRF (Republic of Korea); LAS (Lithuania); MOE and UM (Malaysia); BUAP, CINVESTAV, CONACYT, LNS, SEP, and UASLP-FAI (Mexico); MBIE (New Zealand); PAEC (Pakistan); MSHE and NSC (Poland); FCT (Portugal); JINR (Dubna); MON, RosAtom, RAS and RFBR (Russia); MESTD (Serbia); SEIDI and CPAN (Spain); Swiss Funding Agencies (Switzerland); MST (Taipei); ThEPCenter, IPST, STAR and NSTDA (Thailand); TUBITAK and TAEK (Turkey); NASU and SFFR (Ukraine); STFC (United Kingdom); DOE and NSF (USA).

Individuals have received support from the Marie-Curie program and the European Research Council and EPLANET (European Union); the Leventis Foundation; the A. P. Sloan Foundation; the Alexander von Humboldt Foundation; the Belgian Federal Science Policy Office; the Fonds pour la Formation \`a la Recherche dans l'Industrie et dans l'Agriculture (FRIA-Belgium); the Agentschap voor Innovatie door Wetenschap en Technologie (IWT-Belgium); the Ministry of Education, Youth and Sports (MEYS) of the Czech Republic; the Council of Science and Industrial Research, India; the HOMING PLUS program of the Foundation for Polish Science, cofinanced from European Union, Regional Development Fund, the Mobility Plus program of the Ministry of Science and Higher Education, the OPUS program contract 2014/13/B/ST2/02543 and contract Sonata-bis DEC-2012/07/E/ST2/01406 of the National Science Center (Poland); the Thalis and Aristeia programs cofinanced by EU-ESF and the Greek NSRF; the National Priorities Research Program by Qatar National Research Fund; the Programa Clar\'in-COFUND del Principado de Asturias; the Rachadapisek Sompot Fund for Postdoctoral Fellowship, Chulalongkorn University and the Chulalongkorn Academic into Its 2nd Century Project Advancement Project (Thailand); and the Welch Foundation, contract C-1845.
\end{acknowledgments}

\bibliography{auto_generated}

\cleardoublepage \appendix\section{The CMS Collaboration \label{app:collab}}\begin{sloppypar}\hyphenpenalty=5000\widowpenalty=500\clubpenalty=5000\textbf{Yerevan Physics Institute,  Yerevan,  Armenia}\\*[0pt]
V.~Khachatryan, A.M.~Sirunyan, A.~Tumasyan
\vskip\cmsinstskip
\textbf{Institut f\"{u}r Hochenergiephysik der OeAW,  Wien,  Austria}\\*[0pt]
W.~Adam, E.~Asilar, T.~Bergauer, J.~Brandstetter, E.~Brondolin, M.~Dragicevic, J.~Er\"{o}, M.~Flechl, M.~Friedl, R.~Fr\"{u}hwirth\cmsAuthorMark{1}, V.M.~Ghete, C.~Hartl, N.~H\"{o}rmann, J.~Hrubec, M.~Jeitler\cmsAuthorMark{1}, A.~K\"{o}nig, I.~Kr\"{a}tschmer, D.~Liko, T.~Matsushita, I.~Mikulec, D.~Rabady, N.~Rad, B.~Rahbaran, H.~Rohringer, J.~Schieck\cmsAuthorMark{1}, J.~Strauss, W.~Treberer-Treberspurg, W.~Waltenberger, C.-E.~Wulz\cmsAuthorMark{1}
% ...
\vskip\cmsinstskip
% ...
1:~~Also at Vienna University of Technology, Vienna, Austria\\
% ...

\end{sloppypar}
\end{document}